**Incoherent dielectric tensor tomography for quantitative 3D measurement of biaxial anisotropy**

Juheon Lee[1,2], Yeon Wook Kim[4], Hwanseok Chang[3], Herve Hugonnet[1,2], Seung-Mo Hong[4], Seokwoo Jeon[3], YongKeun Park[1,2,5*]

[1] *Department of Physics, Korea Advanced Institute of Science and Technology (KAIST), Daejeon 34141, South Korea;*

[2] *KAIST Institute for Health Science and Technology, KAIST, Daejeon 34141, South Korea;*

[3] *Department of Materials Science and Engineering, Korea University, Seoul 02841, Republic of Korea*

[4] *Department of Pathology, Asan Medical Center, University of Ulsan College of Medicine, Seoul 05505, Republic of Korea*

[5] *Tomocube Inc., Daejeon 34109, South Korea*

*corresponding authors: Y.K.P (yk.park@kaist.ac.kr)*

**ABSTRACT**

Biaxial anisotropy, arising from distinct optical responses along three principal directions, underlies the complex structure of many crystalline, polymeric, and biological materials. However, existing techniques such as X-ray diffraction and electron microscopy require specialized facilities or destructive preparation and cannot provide full three-dimensional (3D) information. Here we introduce incoherent dielectric tensor tomography (iDTT), a non-interferometric optical imaging method that quantitatively reconstructs the 3D dielectric tensor under incoherent, polarization-diverse illumination. By combining polarization diversity and angular-spectrum modulation, iDTT achieves speckle-free and vibration-robust mapping of biaxial birefringence with submicron resolution. Simulations and experiments on uniaxial and biaxial samples validate its quantitative accuracy. Applied to mixed and polycrystalline materials, iDTT distinguishes crystal types by their birefringent properties and reveals 3D grain orientations and boundaries. This approach establishes iDTT as a practical and accessible tool for quantitative, label-free characterization of biaxial anisotropy in diverse materials.

**INTRODUCTION**

Many functional materials exhibit optical anisotropy, where physical properties vary with direction. The most general form, biaxial anisotropy, describes systems with distinct responses along three orthogonal directions—reflecting complex internal organization in crystalline lattices, polymer chains, or stress-induced microstructures[1–8]. A uniaxial material, by contrast, possesses rotational symmetry around a single optic axis and thus represents a special case of anisotropy. While uniaxial anisotropy can often be measured using conventional optical or crystallographic tools, accurate quantification of biaxial anisotropy remains essential for a complete understanding of anisotropic materials such as low-symmetry crystals, oriented polymers, and fiber-reinforced or strain-engineered media.

X-ray diffraction (XRD) has long been the principal method for quantifying crystallographic anisotropy[9]. By analyzing diffracted intensity patterns, XRD provides statistical information on lattice orientations and texture distributions. Advanced forms such as three-dimensional (3D) X-ray diffraction tomography can even reconstruct individual grain orientations[10–12]. However, achieving spatial resolutions below ~10 µm typically requires synchrotron radiation facilities, limiting accessibility and throughput. Electron microscopy techniques offer higher spatial resolution but with other trade-offs. For instance, electron backscatter diffraction (EBSD) provides quantitative orientation mapping in two dimensions but requires highly polished surfaces and cannot access buried structures[13–16]. Transmission electron microscopy (TEM) can achieve nanometer-scale resolution[17], but is restricted to ultrathin specimens and small fields of view. Consequently, neither X-ray nor electron-based methods are ideal for non-destructive, volumetric analysis of complex anisotropy.

Optical approaches offer a powerful alternative because birefringence—the polarization-dependent refractive index (RI) of light—directly reflects a material's intrinsic anisotropy. Holotomography (HT), which reconstructs 3D RI tomograms, has enabled label-free visualization of cellular and material structures in 3D[18–23]. However, conventional HT treats the RI as a scalar quantity and therefore neglects polarization-dependent information. Polarized light microscopy, using crossed polarizers, has long been used to visualize birefringence[24], yet it provides only two-dimensional (2D) projected information, where uniaxial and biaxial birefringence cannot be distinguished. Conoscopic illumination techniques visualize interference patterns of the optic axes[25,26], but are limited to uniform bulk samples and lack spatial resolution sufficient for microscopic analysis. Thus, a method capable of resolving 3D distributions of anisotropic optical properties is needed.

Recently, dielectric tensor tomography (DTT) was introduced as a quantitative 3D birefringence imaging technique that reconstructs the dielectric tensor, a 3×3 matrix describing the full optical response of anisotropic media[27,28]. In principle, DTT can completely characterize both uniaxial and biaxial anisotropy. However, conventional implementations rely on interferometric measurements, which are highly sensitive to speckle noise and mechanical vibrations. In practice, these instabilities restrict DTT's applicability to simple uniaxial samples. Recent non-interferometric optical methods using incoherent illumination have mitigated some of these challenges[29,30], but their simplifying assumptions—such as in-plane or uniaxial constraints—preclude accurate dielectric tensor tomography reconstruction of biaxial anisotropy.

Here, we introduce incoherent dielectric tensor tomography (iDTT), a non-interferometric optical imaging technique that quantitatively reconstructs the full 3D dielectric tensor under incoherent, polarization-diverse illumination. By extending phase-deconvolution microscopy to a vectorial framework, iDTT retrieves all tensor components through multi-channel deconvolution with optical transfer functions defined by angular-spectrum modulation and polarization diversity. This approach eliminates coherence-related artifacts and enables speckle-free, vibration-robust reconstruction of weak birefringence. The accuracy and versatility of iDTT are verified through numerical simulations and experimental validations on diverse materials, including liquid-crystal particles, uniaxial and biaxial crystals, and biological tissues. Finally, we demonstrate that iDTT can classify mixed crystal systems based on their principal RI differences and resolve 3D grain textures and orientation-dependent boundaries in polycrystalline ulexite, establishing iDTT as a practical route for quantitative optical analysis of biaxial anisotropy.

## Results

**iDTT system**

To achieve quantitative and robust dielectric tensor tomography, we developed iDTT—a non-interferometric, LED-based imaging system that reconstructs the full 3D dielectric tensor under incoherent illumination. By integrating polarization diversity and angular-spectrum modulation, iDTT retrieves all tensor components without coherence artifacts, providing speckle-free and vibration-insensitive measurements of both uniaxial and biaxial anisotropy (Fig. 1). Conceptually, iDTT extends the principle of phase-deconvolution microscopy[31–33] to a vectorial framework, reconstructing the dielectric tensor by deconvolving *z*-stacked intensity data with polarization-dependent optical transfer functions (OTFs).

The optical configuration of iDTT is shown in Fig. 1a. Two orthogonally mounted light-emitting diodes (LEDs) illuminate the sample through a wire-grid polarizing beam splitter (WPBS), enabling rapid switching between orthogonal linear polarization states without mechanical or electronic control. The illumination beam is angularly modulated at the pupil plane by a digital micromirror device (DMD) and subsequently converted to circular polarization using a quarter-wave plate (QWP). Light scattered from the sample passes through an analyzer and is recorded by a polarization camera equipped with a micro-polarizer array, allowing simultaneous acquisition of four polarization-resolved images.

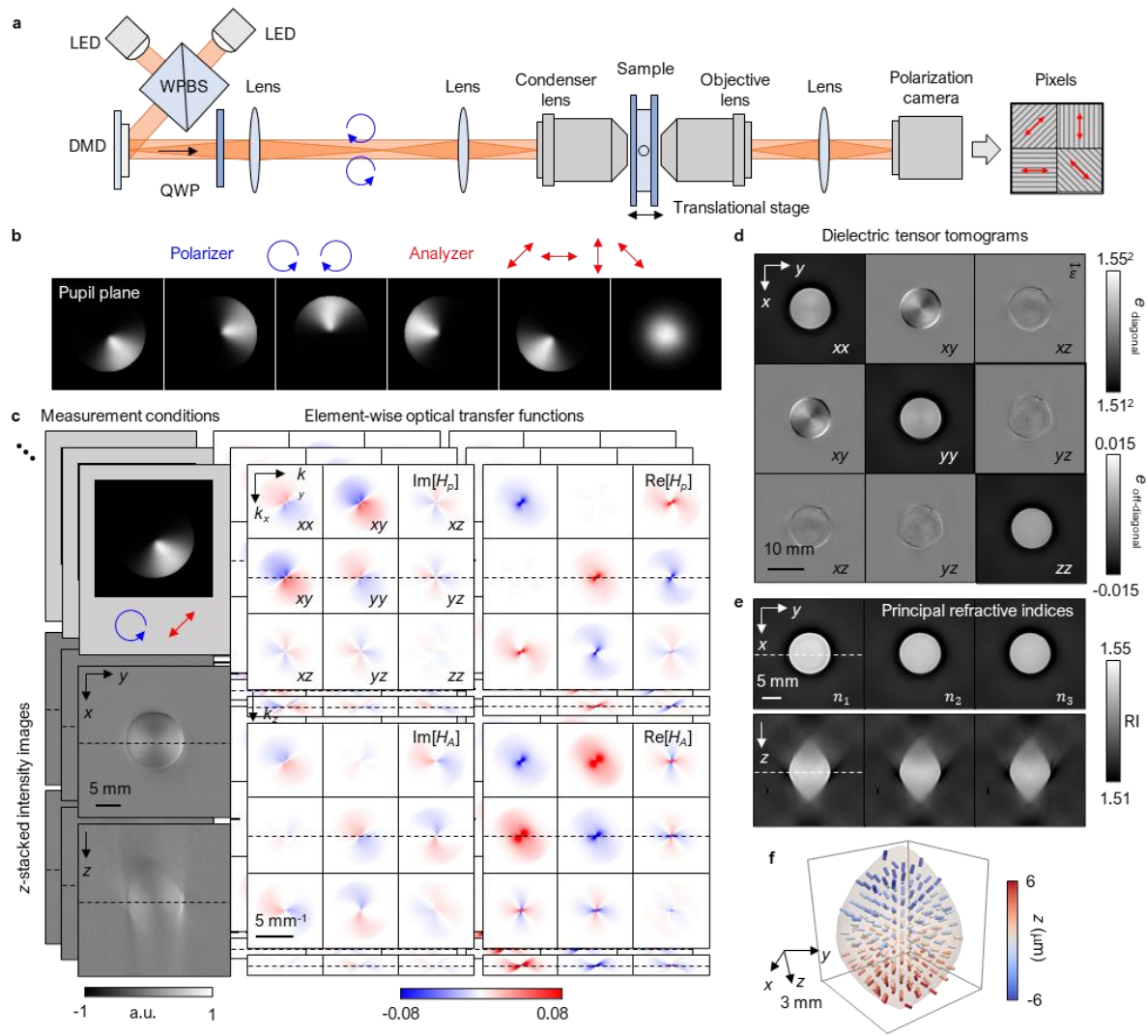

**Figure 1 | Principle and reconstruction pipeline of iDTT. a,** Optical configuration. LED, light-emitting diode; WPBS, wire-grid polarizing beam splitter; DMD, digital micromirror device; QWP, quarter-wave plate;. **b,** Illumination and detection conditions. Six pupil-plane intensity patterns, two circular polarization states for illumination, and four linear polarization directions for detection generate a total of 48 independent OTFs. **c,** Representative measurement set. Each OTF, defined by the illumination–detection pair, produces a stack of polarization-resolved intensity images along *z*. **d,** Reconstruction of the dielectric tensor. Using the measured datasets and the corresponding OTFs, all nine tensor components (real and imaginary parts) are retrieved by inverse deconvolution. **e,** Principal RI tomograms. Diagonalization of the reconstructed tensor yields three spatial maps of principal RIs ($n_1$, $n_2$, $n_3$), representing the magnitude of birefringence along each principal axis. **f,** 3D orientation of optical anisotropy. Rendering of the eigen-axes of the dielectric tensor visualizes the local orientation and relative strength of birefringence throughout the volume.

### iDTT reconstruction algorithm

The reconstruction principle of iDTT is based on the vectorial form of Fourier diffraction theory. When a plane wave sequentially passes through an input polarizer, a birefringent specimen, and an analyzer (Fig. 1a), the unscattered field $u_o$ within the total transmitted field $u = u_o + u_s$ can be expressed as

$$u_o = e^{i\boldsymbol{k_o} \cdot \boldsymbol{r}} \mathbf{a}^*(\boldsymbol{k_o}) \mathbf{p}(\boldsymbol{k_o}), \tag{1}$$

where $\boldsymbol{r}$ is a position vector, $\boldsymbol{k_o}$ is an incident wave vector, $\mathbf{p}(\boldsymbol{k})$ and $\mathbf{a}(\boldsymbol{k})$ denote the polarization states after the polarizer and analyzer, respectively (* indicates the conjugate transpose). This relation describes that an incident plane wave $e^{i\boldsymbol{k_o} \cdot \boldsymbol{r}}$ is first polarized by $\mathbf{p}(\boldsymbol{k_o})$ and then analyzed by $\mathbf{a}(\boldsymbol{k_o})$.

According to the Fourier diffraction theorem[34], the scattered field $u_s$ in the spatial-frequency domain can be written as

$$U_s(\boldsymbol{k}) = \mathrm{FT}[u_s] = \mathbf{a}^*(\boldsymbol{k}) \mathbf{G}(\boldsymbol{k}) \mathbf{F}(\boldsymbol{k} - \boldsymbol{k_o}) \mathbf{p}(\boldsymbol{k_o}) \tag{2}$$

where captical letter denote the Fourier transforms of their corresponding spatial-domain quantities: $\mathbf{G}(\boldsymbol{k})$ and $\mathbf{F}(\boldsymbol{k})$ are the Fourier transforms of the dyadic Green's function tensor $\mathbf{g}(\boldsymbol{r})$ and the scattering potential tensor $\mathbf{f}(\boldsymbol{r}) = k_o^2(\boldsymbol{\varepsilon}(\boldsymbol{r})/n_m^2 - \mathbf{I})$, respectively; $\boldsymbol{\varepsilon}(\boldsymbol{r})$ is the dielectric tensor of a birefringent specimen, $n_m$ is the RI of the surrounding medium, and $\mathbf{I}$ is the identity tensor.

Using Eqs. (1) and (2), the normalized scatterd intensity $s = (|u|^2 - |u_o|^2)/|u_o|^2$ can be expressed in Fourier space as

$$S(\boldsymbol{k}) = \frac{\mathbf{a}^*(\boldsymbol{k}+\boldsymbol{k_o})\mathbf{G}(\boldsymbol{k}+\boldsymbol{k_o})\mathbf{F}(\boldsymbol{k})\mathbf{p}_k(\boldsymbol{k_o})}{[\mathbf{a}^*(\boldsymbol{k_o})\mathbf{p}(\boldsymbol{k_o})]} + c.c. + O(|u_s|^2), \tag{3}$$

where "c.c." denotes the complex conjugate term and $O(|u_s|^2)$ is a square term of $u_s$. Note that the square term can be omitted under weak scattering approximation.

The scattering potential tensor can be decomposed into its real and imaginary part, $f(r) = f_r(r) + if_i(r)$, leading to

$$S(\boldsymbol{k}) = \sum_{m,k}\left[\mathrm{H}_{r,mk}(\boldsymbol{k};\boldsymbol{k_o})\mathrm{F}_{r,mk}(\boldsymbol{k}) + \mathrm{H}_{i,mk}(\boldsymbol{k};\boldsymbol{k_o})\mathrm{F}_{i,mk}(\boldsymbol{k})\right], \tag{4}$$

where $m$ and $k$ are tensor indices, and $\mathrm{H}_{r,mk}$ and $\mathrm{H}_{i,mk}$ denote the element-wise OTFs for the real and imaginary parts of the dielectric tensor. Their explicit forms are

$$\mathrm{H}_{r,mk}(\boldsymbol{k};\boldsymbol{k_o}) = \sum_n \left[\frac{a_n^*(\boldsymbol{k}+\boldsymbol{k_o})G_{nm}(\boldsymbol{k}+\boldsymbol{k_o})\mathrm{p}_k(\boldsymbol{k_o})}{\mathbf{a}^*(\boldsymbol{k_o})\mathbf{p}(\boldsymbol{k_o})} + \frac{a_n(-\boldsymbol{k}+\boldsymbol{k_o})G_{nm}^*(-\boldsymbol{k}+\boldsymbol{k_o})\mathrm{p}_k^*(\boldsymbol{k_o})}{\mathbf{p}^*(\boldsymbol{k_o})\mathbf{a}(\boldsymbol{k_o})}\right], \tag{5}$$

$$\mathrm{H}_{i,mk}(\boldsymbol{k};\boldsymbol{k_o}) = \sum_n i\left[\frac{a_n^*(\boldsymbol{k}+\boldsymbol{k_o})G_{nm}(\boldsymbol{k}+\boldsymbol{k_o})\mathrm{p}_k(\boldsymbol{k_o})}{\mathbf{a}^*(\boldsymbol{k_o})\mathbf{p}(\boldsymbol{k_o})} - \frac{a_n(-\boldsymbol{k}+\boldsymbol{k_o})G_{nm}^*(-\boldsymbol{k}+\boldsymbol{k_o})\mathrm{p}_k^*(\boldsymbol{k_o})}{\mathbf{p}^*(\boldsymbol{k_o})\mathbf{a}(\boldsymbol{k_o})}\right]. \tag{6}$$

where $n$ is a tensor index. Unlike interferometric DTT, iDTT employs incoherent illumination, which can be described as an incoherent superposition of multiple plane waves whose intensities are modulated at the pupil plane. When the angular intensity distribution at the pupil is denoted as $\rho(\boldsymbol{k_o})$, the effective OTFs become

$$\mathbf{H}_r(\boldsymbol{k}) = \int \rho(\boldsymbol{k_o})\mathbf{H}_r(\boldsymbol{k};\boldsymbol{k_o})d\boldsymbol{k_o}, \tag{7}$$

$$\mathbf{H}_i(\boldsymbol{k}) = \int \rho(\boldsymbol{k_o})\mathbf{H}_i(\boldsymbol{k};\boldsymbol{k_o})d\boldsymbol{k_o}. \tag{8}$$

To solve all elements of the dielectric tensor, multiple measurements with distinct OTFs are required. In practice, the OTFs are diversified by modulating the pupil intensity, input polarization, and detection polarization states.

Six pupil patterns, two circular illumination polarizations, and four analyzer orientations were used (Fig. 1b), generating a set of 48 independent measurements. At each spatial frequency, the data can be represented in matrix form as

$$\mathbf{y} = \mathbf{A}\mathbf{x} = \begin{bmatrix} S_1 \\ S_2 \\ \vdots \end{bmatrix} = \begin{bmatrix} H_{r,11,1} & H_{r,22,1} & \cdots & H_{i,11,1} & H_{i,22,1} & \cdots \\ H_{r,11,2} & H_{r,22,2} & \cdots & H_{i,11,2} & H_{i,22,2} & \cdots \\ \vdots & \vdots & & & & \ddots \end{bmatrix} \begin{bmatrix} F_{r,11} \\ F_{r,22} \\ \vdots \\ F_{i,11} \\ F_{i,22} \\ \vdots \end{bmatrix}, \qquad (9)$$

where $\mathbf{y}$ contains the measured intensities and x represent the unknown tensor elements, $\mathbf{x} = [F_{r,11}, F_{r,22}, F_{r,33}, F_{r,12}, F_{r,13}, F_{r,23}, F_{i,11}, F_{i,22}, F_{i,33}, F_{i,12}, F_{i,13}, F_{i,23}]^T$. Since the dielectric tensor is symmetric, only six independent components exist for each of the real and imaginary parts. The solution $x$ is obtained by a regularized pseudoinverse of $\mathbf{A}$, as detailed in *Methods*. Finally, the inverse Fourier transform of $\mathbf{x}$ yields the 3D spatial distribution of all dielectric-tensor components (Fig. 1d).

**Diagonalization of the dielectric tensor**

The local tensor diagonalization yields both the magnitude of birefringence and the 3D orientation of anisotropy, enabling quantitative visualization of uniaxial and biaxial optical properties. To extract physically meaningful quantities from the reconstructed tensor, the dielectric tensor $\boldsymbol{\varepsilon}(\boldsymbol{r})$ is diagonalized at each voxel as

$$\boldsymbol{\varepsilon} = \mathbf{R} \begin{bmatrix} n_1^2 & 0 & 0 \\ 0 & n_2^2 & 0 \\ 0 & 0 & n_3^2 \end{bmatrix} \mathbf{R}^T, \qquad (10)$$

where $n_1$, $n_2$, and $n_3$ are the square roots of the eigenvalues ($n_1 > n_2 > n_3$), representing the principal RIs, and $\mathbf{R}$ is the rotation matrix whose columns are the corresponding eigenvectors, indicating the local orientation of the principal optical axes (Fig. 1e,f).

The magnitudes and orientations of the principal RIs describe the strength and directional characteristics of optical anisotropy. For a uniaxial medium, two of the principal RIs are equal, and the remaining distinct RI defines the optic axis: $n_1 > n_2 = n_3$ for a positive uniaxial material and $n_1 = n_2 > n_3$ for a negative uniaxial material. In contrast, a biaxial medium exhibits three unequal principal RIs ($n_1 > n_2 > n_3$), corresponding to three mutually orthogonal principal axes.

**Numerical simulation of iDTT reconstruction**

To verify the accuracy and robustness of the proposed reconstruction principle, we performed numerical simulations in which virtual anisotropic samples were forward-modeled and then reconstructed using iDTT (Fig. 2).

Two representative cases were considered: a uniaxial bead with a radial optic-axis configuration and biaxial beads with distinct principal-axis orientations. Forward simulations of the polarization-resolved intensity images were generated using the convergent Born series method, which accurately models light propagation in anisotropic media[35,36]. For each predefined combination of illumination and detection conditions, multiple intensity images were computed and used as input data for tensor reconstruction. After reconstruction, the resulting dielectric tensors were diagonalized to obtain the tomograms of the principal RIs and their corresponding orientations.

The reconstructed tomograms were compared with the ground-truth models to assess the reconstruction accuracy (Figs. 2a–f). The reconstructed results closely reproduced the missing-cone ground truths for both uniaxial and biaxial samples, confirming that iDTT reliably recovers the 3D dielectric-tensor distributions under realistic imaging conditions. In the orientation maps, hue encodes the local direction of the principal axis and brightness represents the birefringence magnitude ($\Delta n = n_1 - n_2$); desaturated colors indicate directions tilted out of the image plane.

To further examine the quantitative accuracy of the recovered RIs, line profiles of the principal RIs were extracted from the biaxial sample (Fig. 2g–i). Compared with the ideal ground truth, the missing-cone ground truth shows a small reduction in RI magnitude, as expected from its limited angular sampling. The reconstructed profiles exhibit a marginally larger underestimation, which cannot be explained by the missing-cone effect alone. This deviation originates from the low-frequency attenuation of the system's incoherent OTFs, which slightly suppresses the absolute RI amplitudes while preserving their relative differences.

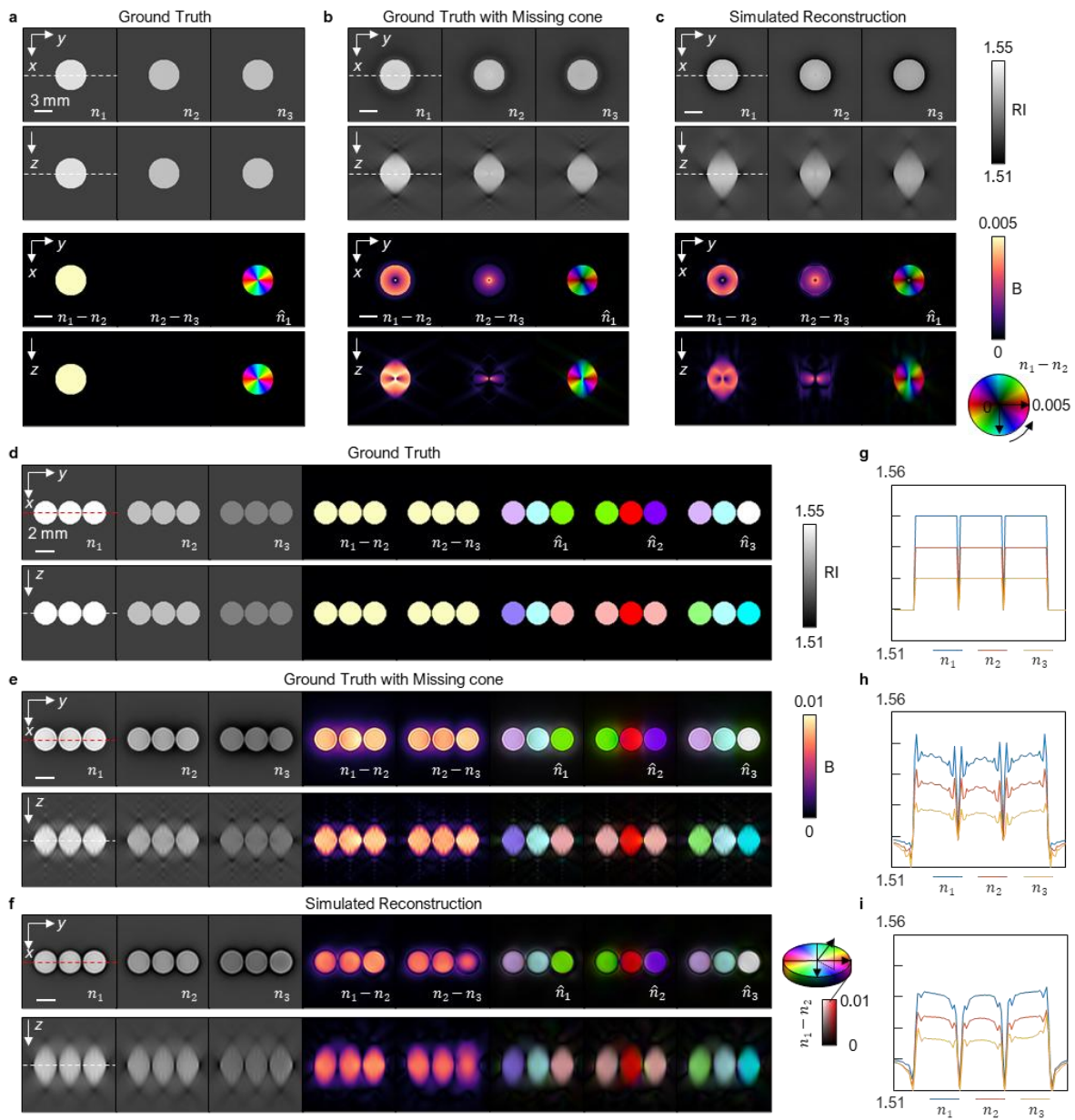

**Figure 2 | Numerical validation of the iDTT reconstruction algorithm. a–c,** Principal RI and orientation tomograms of a positive uniaxial bead with radial alignment: (**a**) ground-truth model, (**b**) ground truth after applying the missing-cone filter, and (**c**) reconstruction from simulated iDTT data. **d–f,** Corresponding results for biaxial beads with distinct principal-axis orientations. In the orientation maps, the hue encodes the local direction of the principal axis and the brightness indicates the birefringence magnitude ($\Delta n = n_1 - n_2$); desaturated colors represent directions tilted out of the image plane. **g–i,** Line profiles of the reconstructed principal RIs ($n_1$, $n_2$, $n_3$) for the biaxial sample, extracted along the red dashed lines in the $n_1$ tomograms of (**d–f**). The reconstructed values closely follow the ground truth with minor underestimation caused by limited low-frequency gain in the system OTFs

**Experimental reconstruction of alignment configuration within liquid crystal particles**

To experimentally validate the proposed iDTT framework, we applied the method to liquid-crystal (LC) microparticles whose internal director configurations are well established. The particles were fabricated from a reactive mesogen mixture, following the same procedure described in our previous DTT study[27,37–39]. Their alignment modes were controlled to form either radial or bipolar configurations by selecting the appropriate surfactant during emulsification. Following measurement, the dielectric tensors were reconstructed using iDTT and subsequently diagonalized to obtain the principal RIs and 3D orientation fields.

The reconstructed tomograms clearly reveal the birefringent structures of the LC particles (Fig. 3a,c). The $n_1$ tomograms show a strong spatial modulation corresponding to the optical axis distribution, while $n_2$ and $n_3$ rremain nearly identical, consistent with positive uniaxial birefringence. This behavior reflects the linear molecular alignment typical of uniaxially symmetric LC phases. The orientation maps of the $n_1$ axis visualize the internal director arrangement within each particle, faithfully reproducing the expected radial and bipolar configurations. These results are further corroborated by the 3D rendered orientation fields (Fig. 3b,d), which exhibit excellent agreement with the known director geometries, demonstrating that iDTT accurately resolves uniaxial alignment structures in anisotropic soft-matter systems.

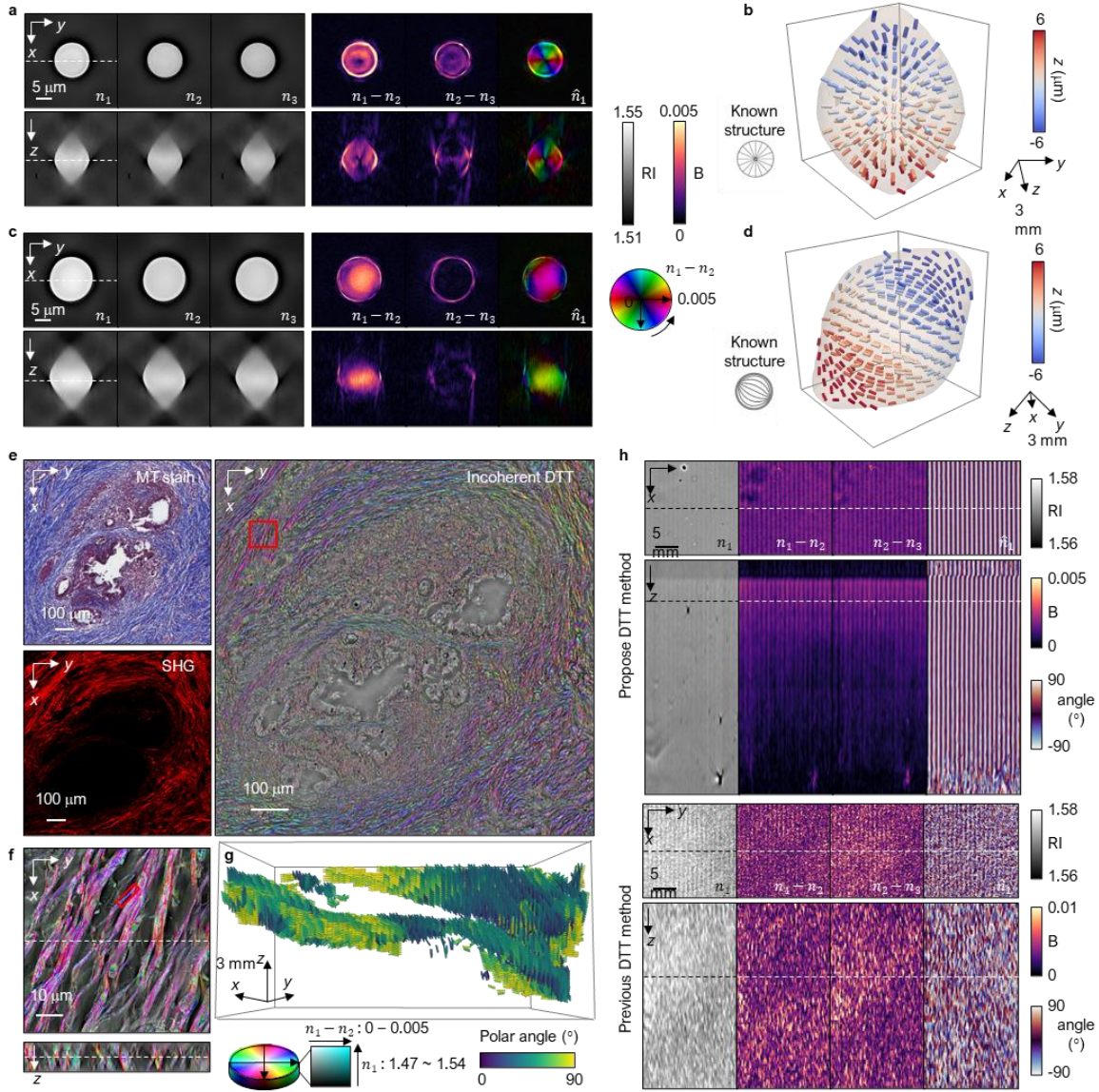

**Figure 3 | Experimental validation of iDTT. a,b,** Principal RI tomograms and orientation maps of LC particles with radial alignment, and the corresponding 3D rendering of the $n_1$-axis orientations compared with the known radial configuration. **c,d,** Results for LC particles with bipolar alignment and the corresponding 3D rendering, confirming that the reconstructed $n_1$-axes reproduce the expected bipolar director field. **e,** Collagen-rich human colon tissue imaged by Masson's trichrome (MT) staining, second-harmonic-generation (SHG) microscopy, and iDTT. The iDTT image reveals birefringent fiber bundles that spatially coincide with collagen structures in the reference modalities. **f,** Magnified region (red box in e) showing that the local birefringence orientations follow the collagen-fiber morphology in the *xy* plane. **g,** Volumetric rendering of the same region, color-coded by polar angle, demonstrates that iDTT quantitatively retrieves the 3D fiber orientation without labeling. **h,** Comparison between conventional interferometric DTT and iDTT for a polarization-volume-grating (PVG) sample. iDTT clearly resolves the periodic molecular-orientation pattern and depth profile while eliminating speckle noise and mechanical-vibration artifacts present in conventional DTT.

**Imaging collagen fiber organization in human colon tissue**

To further validate the capability of iDTT for biological specimens, we applied the method to human colon tissue obtained after radiation therapy, which exhibits dense collagen networks due to radiation-induced fibrosis. The samples were examined using three complementary imaging modalities: Masson's trichrome (MT) staining, second-harmonic generation (SHG) microscopy, and iDTT. Because histochemical staining interferes with label-free optical measurements, serial tissue sections were prepared—one section was stained for MT imaging, while the adjacent unstained section was used for SHG and iDTT. iDTT imaging was conducted over an area of approximately 1 mm × 1 mm by stitching 20 × 20 tomograms, allowing direct comparison of fiber morphology across modalities.

The MT-stained, SHG, and iDTT images are shown side-by-side in Fig. 3e. In the MT image, collagen fibers appear blue, whereas the SHG image highlights optical signals from collagen's non-centrosymmetric structure. Both modalities provide morphological or molecular contrast for identifying collagen-rich regions. In the iDTT results, regions exhibiting strong birefringence correspond precisely to the collagen-dense areas identified by MT and SHG, confirming that the birefringence measured by iDTT originates predominantly from ordered collagen fibers.

A magnified view of a representative region (Fig. 3f) reveals that the birefringence orientations retrieved by iDTT follow the structural direction of collagen bundles in the *xy*-plane. Further zooming into an individual fiber, volumetric rendering of the $n_1$-axis distribution (Fig. 3g) demonstrates that the reconstructed 3D optical-axis orientations coincide with collagen fiber morphology through the tissue thickness, validating the accuracy of iDTT for resolving collagen architecture.

Beyond confirming consistency with MT and SHG, iDTT offers several distinct advantages: it quantitatively maps both birefringence magnitude and 3D orientation, which MT staining cannot provide; it operates under broadband incoherent illumination without requiring femtosecond pulsed lasers or raster scanning as in SHG microscopy; and it enables wide-field, volumetric, label-free imaging with submicron spatial resolution. These advantages establish iDTT as a practical and accessible tool for label-free assessment of fibrotic and collagen-rich tissues.

**Experimental validation of iDTT improvement over conventional DTT**

To assess the performance improvement achieved by incoherent illumination, we compared iDTT with conventional interferometric DTT using a polarization volume grating (PVG) sample previously characterized by the coherent method[40]. The PVG was fabricated from an azobenzene–LC polymer, in which the interference of two orthogonally polarized beams induces a spatially periodic rotation of the molecular orientation with a period of approximately 1 μm, forming a volumetric birefringent grating[41]. The iDTT results were directly compared with the conventional DTT reconstructions to evaluate the improvement in sensitivity and image quality.

Figure 3h presents the comparison of reconstructed principal RIs and $n_1$-axis orientation tomograms obtained by both methods. In the iDTT reconstruction, the periodic molecular-alignment pattern of the grating is clearly resolved, whereas in conventional DTT it is barely visible due to severe speckle noise. In previous studies, recovering this periodic structure required additional filtering and post-processing[42], which limited the ability to identify defects or irregularities in the grating pattern. Moreover, in the *yz* cross-sectional views, conventional DTT suffers from noise-induced artifacts that obscure the thickness and depth profile of the grating layer, while iDTT clearly delineates the depth range over which the periodic orientation is inscribed.

These results demonstrate that the use of incoherent, polarization-diverse illumination effectively suppresses speckle and coherence artifacts, yielding a substantial enhancement in reconstruction fidelity and sensitivity. By eliminating the need for interferometric stability and post-processing, iDTT provides a more reliable and quantitative approach for characterizing 3D birefringent structures in both engineered optical elements and naturally anisotropic materials.

**Measurement of Crystal Samples for Accuracy Verification**

To quantitatively verify the accuracy of iDTT, we measured three crystalline samples with well-characterized principal RIs: α-quartz (BCR070, BCR), cristobalite (NIST1879B, NIST), and calcium chloride (CaCl$_2$) (499609, Sigma-Aldrich). α-quartz is a positive uniaxial crystal[43] ($n_1 = 1.552$, $n_2 = n_3 = 1.543$), cristobalite is negative uniaxial[44] ($n_1 = n_2 = 1.487$, $n_3 = 1.484$), and CaCl$_2$ is biaxial, having three distinct principal RIs[44] ($n_1 = 1.613$, $n_2 = 1.605$, $n_3 = 1.600$). To minimize refraction mismatch and surface scattering, all samples were immersed in index-matching oils with RIs close to those of the crystals. The dielectric tensors were reconstructed and diagonalized to obtain 3D tomograms of the principal RIs and their orientations.

Cross-sectional tomograms of the reconstructed RIs are shown in Fig. 4a–c. The measured birefringence characteristics of each crystal are consistent with their known optical anisotropy. To quantitatively evaluate the reconstruction accuracy, line profiles of the principal RIs were extracted and compared with the reference values (Fig. 4d). The reconstructed absolute RI values are slightly lower than the references—consistent with simulation results—due to the finite transfer gain of the system OTFs at low spatial frequencies. However, the differences between the principal RIs, which correspond to birefringence magnitudes, agree well with the reference data, demonstrating that iDTT reliably preserves relative anisotropy contrast. In CaCl$_2$, small spatial variations in the reconstructed principal RIs are observed, likely caused by rapid local changes in principal-axis orientation within the crystal lattice.

The orientation tomograms further distinguish each crystal grain by its unique optical-axis distribution (Figs. 4a–c). In particular, CaCl$_2$ exhibits lamellar orientation patterns where the optic-axis direction alternates periodically across layers—consistent with its known twinning structure[45]. These lamellar features are less evident in the $n_1$-orientation map but become pronounced in the $n_2$ and $n_3$ orientations. This observation underscores the importance of full-tensor reconstruction in resolving complex biaxial anisotropy, which cannot be accurately characterized by scalar or uniaxial assumptions.

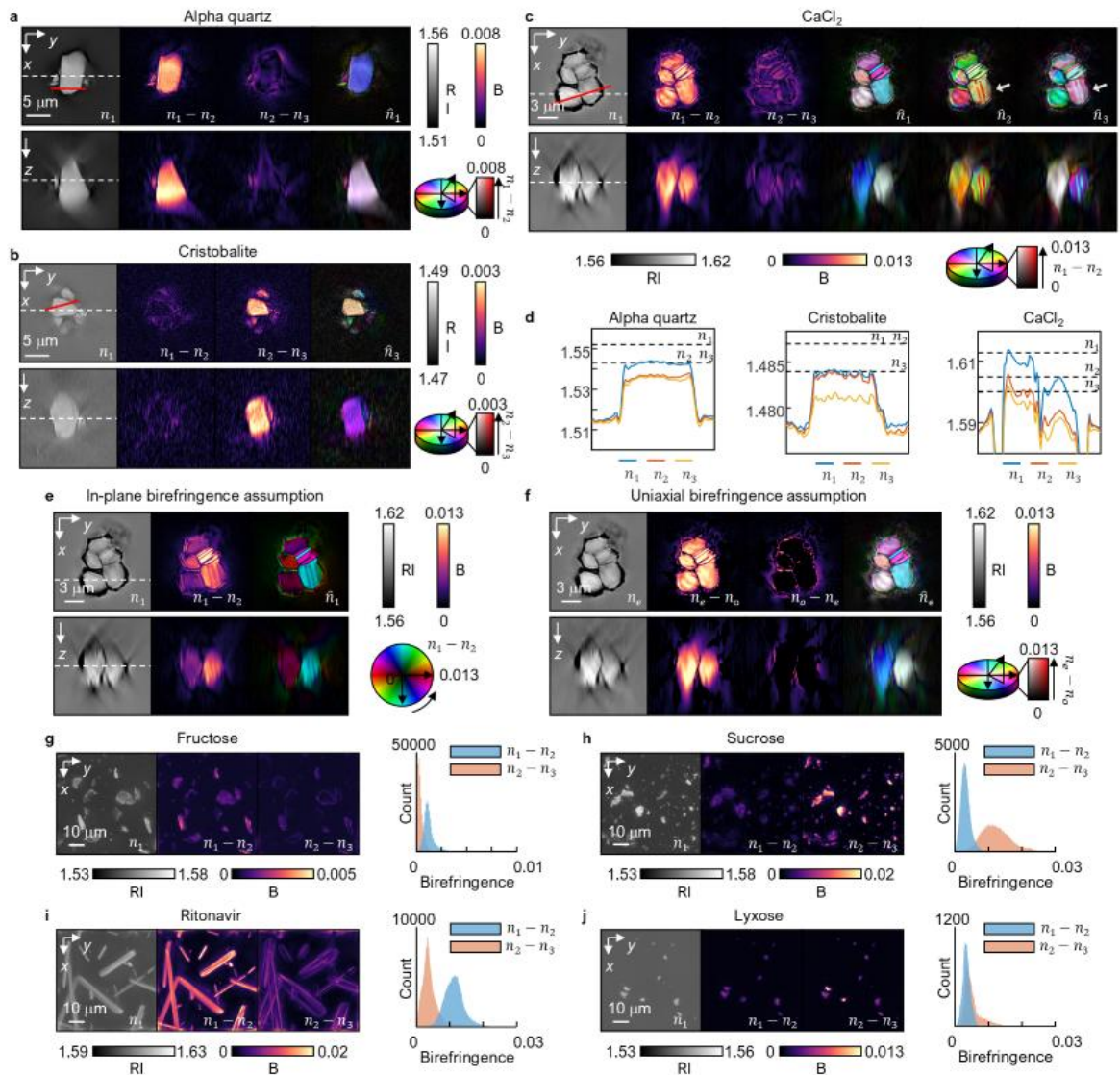

**Figure 4 | Experimental reconstruction and characterization of crystalline samples using iDTT. a–c,** Cross-sectional tomograms of principal RIs ($n_1$, $n_2$, $n_3$), birefringence maps ($n_1 - n_2$, $n_2 - n_3$), and principal orientations for (a) α-quartz, (b) cristobalite, and (c) calcium chloride ($CaCl_2$). White arrows in (c) indicate the lamellar structures within $CaCl_2$. **d,** Line profiles of the reconstructed principal RIs extracted along the red dashed lines in (a–c). Dashed horizontal lines denote the reference values, showing that iDTT quantitatively recovers the birefringence magnitudes while slightly underestimating the absolute RIs due to limited low-frequency gain. **e,f,** Results obtained when constraining the tensor reconstruction with (e) an in-plane birefringence assumption (2 × 2 tensor) or (f) a uniaxial assumption (single optic axis). Both simplified models fail to reproduce the lamellar orientation pattern observed in the full-tensor reconstruction of $CaCl_2$, underscoring the need for complete tensor recovery. **g–j,**

Maximum-intensity projections of reconstructed tomograms and corresponding histograms of RI differences ($n_1 - n_2$, $n_2 - n_3$) for various organic crystals: fructose (g), sucrose (h), ritonavir (i), and lyxose (j). Distinct distributions of RI differences reveal characteristic anisotropic behaviors for each crystal, enabling quantitative classification of uniaxial and biaxial optical properties.

**Necessity of full-tensor reconstruction for biaxial crystals**

To elucidate the importance of full-tensor reconstruction, we compared the measured dielectric tensor of $CaCl_2$ with and without anisotropy constraints. Two representative simplifications commonly adopted in conventional birefringence imaging were tested. The first is the in-plane birefringence assumption, which models a $2 \times 2$ tensor by neglecting all $z$-components[30,46]. The second is the uniaxial assumption, which constrains birefringence to a single optic axis[29,47]. Each constraint was applied to the experimentally measured tensor of $CaCl_2$, and the resulting tensors were diagonalized and compared with the unconstrained full-tensor reconstruction.

The reconstructed principal RIs and orientation tomograms under the two assumptions reveal distinct artifacts (Fig. 4e,f). With the in-plane constraint, the tomogram exhibits orientation-dependent inhomogeneity in the principal RI values and fails to resolve the characteristic lamellar orientation pattern. Under the uniaxial constraint, the reconstruction appears spatially uniform but cannot reproduce three distinct principal RIs or recover the alternating orientation domains. In contrast, the full-tensor reconstruction faithfully resolves all three principal RIs and their orthogonal orientations, clearly depicting the lamellar twinning structure of $CaCl_2$.

**Characterization of Anisotropic Properties in Various Crystals**

To demonstrate the versatility of iDTT for analyzing anisotropic materials, we applied the method to several organic crystalline samples—fructose (1286504, USP), sucrose (NIST17G, NIST), ritonavir (R180000, Simson), and lyxose (220477, Sigma-Aldrich). For each sample, the 3D distributions of the principal RIs were reconstructed and analyzed to determine their birefringent characteristics.

To visualize the spatial and statistical distributions of anisotropy, maximum-intensity projections of the reconstructed RIs and their differences ($n_1 - n_2$, $n_2 - n_3$) are presented along with corresponding histograms (Figs. 4g–j). The voxel values used for the histograms were extracted from crystal regions identified by thresholding the $n_1$ tomograms. The distinct shapes of the histograms reveal characteristic birefringent behaviors for each material: fructose exhibits weak uniaxial birefringence; sucrose shows a larger $n_2 - n_3$ than $n_1 - n_2$; ritonavir displays dominant $n_1 - n_2$ anisotropy; and lyxose exhibits nearly equal values of $n_1 - n_2$ and $n_2 - n_3$, indicative of near-biaxial symmetry.

These results demonstrate that the statistical distributions of RI differences provide a quantitative fingerprint of crystal anisotropy, enabling rapid differentiation of uniaxial and biaxial optical behaviors. Minor RI variations observed among neighboring crystals—particularly in fructose, sucrose, and lyxose—are attributed to slight lattice distortions and conformational flexibility of organic molecular bonds, which locally perturb birefringence.

**Classification of Crystal Types Based on Principal RI Differences**

To further demonstrate the quantitative capability of iDTT, we analyzed a mixture of crystalline particles with distinct birefringent properties. The sample consisted of a random mixture of α-quartz, cristobalite, and ulexite (Suk-Bo Korea; Boron, California, USA). *α*-quartz and cristobalite were the same uniaxial samples shown in Fig. 4a,b, whereas ulexite is a biaxial crystal with three principal RIs[45] ($n_1 = 1.520$, $n_2 = 1.504$, and $n_3 = 1.491$). Each component was reconstructed within the same volumetric dataset, and the individual crystal types were identified based on the differences among their principal RI tomograms.

Figure 5a shows cross-sectional tomograms of the reconstructed $n_1$, $n_1 - n_2$, and $n_2 - n_3$ distributions at several axial depths. From these volumes, crystal-classification maps were generated by applying threshold ranges to the principal RI differences (Fig. 5b). The corresponding threshold values and volume ratios of each crystal type were determined by voxel counting within the classified regions (Fig. 5c). This approach enables label-free quantitative discrimination of multiple crystalline species within a single field of view.

These results highlight the value of iDTT as an effective and accessible tool for quantitative characterization of crystal anisotropy, even for micron-sized or randomly oriented particles. Unlike X-ray diffraction, which requires synchrotron radiation for comparable spatial resolution, or EBSD, which is limited to polished 2D surfaces, iDTT provides high-resolution, volumetric, and non-destructive measurements of 3D optical anisotropy without specialized facilities or sample preparation. This capability establishes iDTT as a powerful complementary method for identifying and classifying crystalline materials through their intrinsic birefringence signatures.

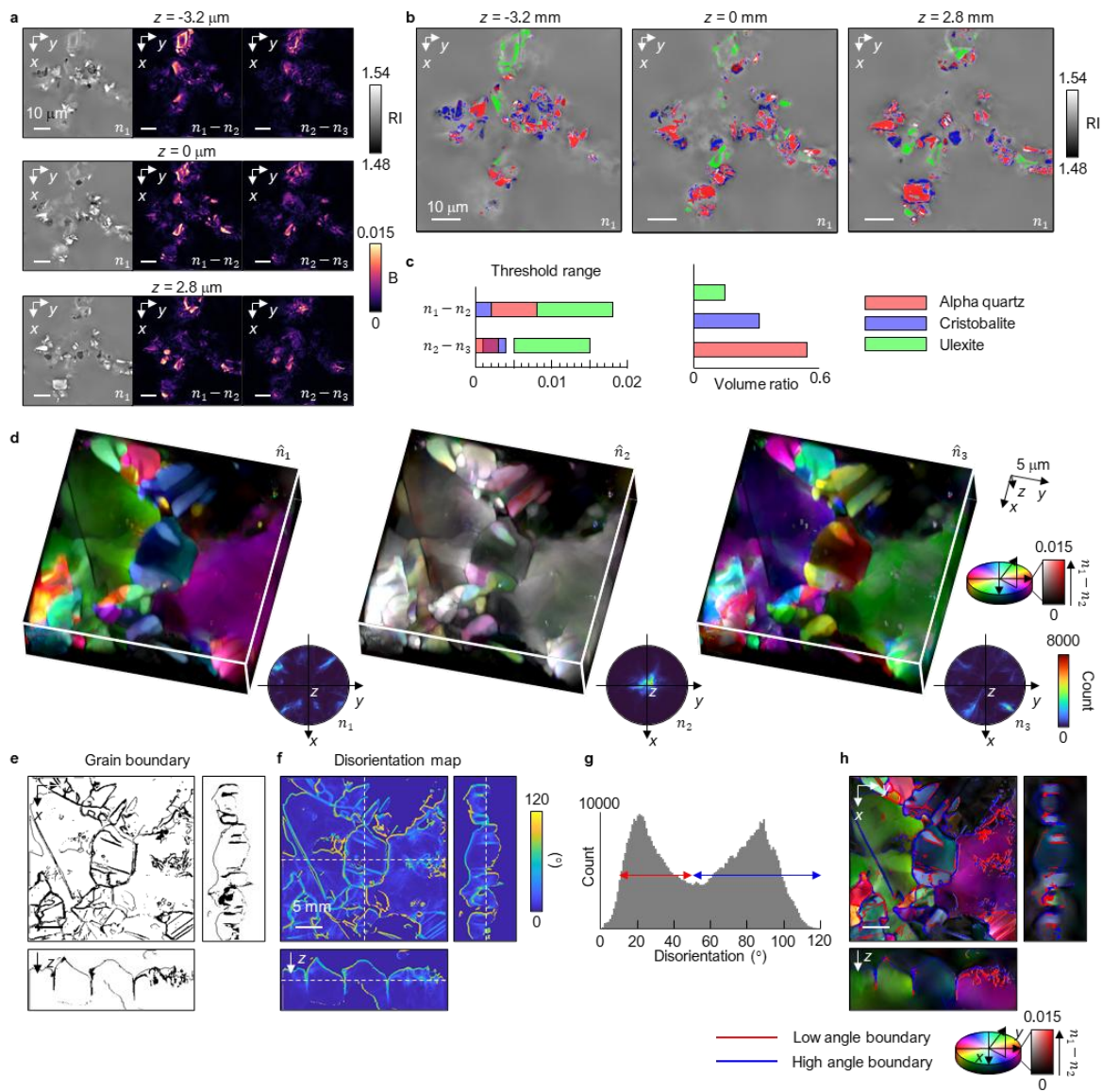

**Figure 5 | Demonstration of iDTT applicability for quantitative anisotropy analysis in mixed and polycrystalline materials. a–c,** Quantitative classification of mixed crystals containing α-quartz, cristobalite, and ulexite. (**a**) Cross-sectional tomograms of $n_1$, $n_1 - n_2$, and $n_2 - n_3$ at multiple axial planes. (**b**) Crystal-type classification maps obtained by applying threshold ranges of principal RI differences to the reconstructed tomograms. (**c**) Threshold ranges used for each crystal type and the corresponding volume ratios estimated from the classification masks. **d–h,** Characterization of microstructural anisotropy in polycrystalline ulexite. (**d**) 3D rendered volume of principal orientations, with pole figures showing the orientation distributions of the principal axes. (**e**) Grain boundaries identified from local orientation discontinuities. (**f**) Map of disorientation angles between neighboring voxels, quantifying orientation inhomogeneity. (**g**) Histogram of disorientation angles at grain boundaries, revealing two dominant regimes corresponding to small (red) and large (blue) orientation differences. (**h**) Grain boundaries associated with these two regimes are overlaid on the $n_1$ orientation tomogram, highlighting distinct inter-grain connectivity and misorientation characteristics.

**3D Orientation and Grain Boundary Analysis of Polycrystalline Material**

To further demonstrate the applicability of iDTT for structural analysis, we investigated a polycrystalline ulexite (Suk-Bo Korea; Boron, California, USA) sample composed of multiple grains with distinct orientations. Ulexite naturally forms needle-shaped grains exhibiting a preferred morphological alignment along a common crystallographic direction[48]. A bulk crystal was mounted on a cover glass with its needle axis oriented perpendicular to the substrate and then mechanically polished into a thin section using graded abrasives. Following iDTT measurement, the reconstructed dielectric tensors were diagonalized to obtain the 3D distributions of the principal optical axes.

The 3D rendered orientation maps clearly resolve individual grains according to their principal-axis orientations (Fig. 5d).To further illustrate the polycrystalline texture, orientation distributions for each principal axis were plotted as pole figures. Among them, the $n_2$ axis shows a predominant alignment along the $z$-direction, consistent with the needle axis of ulexite grains and in agreement with previously reported anisotropy characteristics[48].

Based on the reconstructed orientation fields, we quantified the disorientation of optical anisotropy, which represents the local angular mismatch between neighboring voxels (see *Methods* for the mathematical definition). The computed disorientation maps (Fig. 5e,f) reveal clear grain boundaries, where orientation discontinuities exceed a defined threshold. The histogram of disorientation angles (Fig. 5g) shows two distinct populations corresponding to small and large angular differences. Regions with small disorientation angles represent coherent grain boundaries with high lattice continuity, whereas those with large angular differences correspond to incoherent boundaries associated with weaker inter-grain coupling. By distinguishing these two boundary types, iDTT enables inference of the microstructural integrity and mechanical heterogeneity within the polycrystalline sample (Fig. 5h).

These results demonstrate the advantages of iDTT as a non-destructive 3D method for mapping anisotropy and orientation in polycrystalline materials, in contrast to EBSD, which is limited to 2D characterization and requires destructive sectioning for volumetric analysis.

**Conclusion**

We developed iDTT, a quantitative optical imaging method that reconstructs all components of the 3D dielectric tensor under incoherent, polarization-diverse illumination. By deconvolving measured intensity data with polarization-dependent OTFs, iDTT retrieves the principal RIs and their 3D orientations, enabling direct, quantitative mapping of uniaxial and biaxial anisotropy in complex materials.

Through comprehensive numerical and experimental validation, we demonstrated that iDTT accurately reconstructs biaxial anisotropy across a wide range of systems. Simulations verified quantitative tensor recovery for both uniaxial and biaxial models, while experiments with LC particles, crystalline materials, and biological tissues confirmed reliable performance on real specimens. The method successfully distinguished multiple crystal types in mixed systems and resolved the 3D grain textures and orientation-dependent boundaries in polycrystalline ulexite, highlighting its capability for volumetric analysis of complex optical anisotropy.

Compared with established techniques, iDTT offers a practical, high-resolution, and non-destructive alternative for anisotropy analysis at the micro- to nanoscale. Unlike X-ray diffraction, which requires synchrotron facilities for similar spatial resolution, or EBSD, which is limited to 2D and destructive measurements, iDTT exploits intrinsic optical properties to provide label-free, volumetric characterization of anisotropy in a simple optical setup.

Furthermore, by eliminating the interferometric configuration required in conventional DTT, iDTT effectively suppresses speckle noise and mechanical instability, achieving robust and quantitative reconstruction under incoherent illumination. Compared with recently proposed incoherent tensor-imaging frameworks[29,30], iDTT uniquely reconstructs all six independent components of the symmetric dielectric tensor in real and imaginary domains, whereas previous methods primarily estimated effective anisotropy magnitudes. The experimental validation across uniaxial, biaxial, and biological samples demonstrates broader applicability and quantitative robustness beyond prior implementations. This innovation extends the practical applicability of dielectric-tensor tomography from uniaxial to biaxial materials, enabling optical access to weak or spatially varying birefringence that was previously inaccessible.

Despite its demonstrated capability, iDTT also presents several technical limitations that warrant further improvement. First, the absolute RIs reconstructed by iDTT tend to be slightly underestimated due to the limited low-frequency transfer gain of incoherent OTF. This can be compensated through calibration using isotropic reference samples or by incorporating iterative reconstruction algorithms that jointly recover the missing DC components. Second, while the current system achieves submicron resolution, the axial resolution and sensitivity to weak birefringence ($<10^{-5}$) remain constrained by the numerical aperture and photon budget of LED illumination. Enhancing illumination angular bandwidth or employing high-dynamic-range detection schemes could further extend the measurable anisotropy range. Lastly, the present reconstruction assumes a single-scattering regime, which may not fully capture multiple-scattering effects in highly turbid or thick specimens. Adopting nonlinear forward models or learning-based regularization frameworks could help generalize iDTT to such complex optical environments[49–51]. In addition, because the reconstruction framework is wavelength independent, iDTT can, in principle, be extended beyond the visible range—enabling implementations in the infrared, ultraviolet, or even X-ray spectral domains—thereby broadening its applicability to diverse materials and imaging contexts[52,53].

Overall, iDTT establishes a new optical route for the direct, quantitative, and volumetric measurement of biaxial anisotropy. By bridging crystallographic metrology and optical imaging, it provides an accessible and versatile platform for investigating anisotropic structures in diverse systems—from inorganic crystals to polymers and biological tissues—opening new opportunities for non-destructive 3D analysis of complex anisotropy.

**Methods**

**Experimental iDTT setup**

The proposed iDTT system employs a non-interferometric optical configuration operating under incoherent illumination (Fig. 1a). The illumination module consists of two LEDs (center wavelength = 625 nm, bandwidth = 17 nm; M625L4, Thorlabs) mounted on the transmissive and reflective sides of a wire-grid polarizing beam splitter (WPBS). By alternately activating the two LEDs, the incident polarization state is switched between orthogonal linear polarizations without mechanical rotation of optical elements.

The polarized beam transmitted through the WPBS is directed to a DMD (Vialux V-7001 VIS) positioned at a conjugate pupil plane. The DMD projects grayscale patterns that spatially modulate the angular intensity distribution of illumination. After reflection, the beam passes through a quarter-wave plate (QWP) to convert the linear polarization into circular polarization. Through a 4f relay system (focal lengths = 400 mm and 200 mm), the circularly polarized light is imaged onto the pupil plane of a condenser lens (UPLFLN100XOP-2, Olympus).

Light scattered from the sample is collected by an objective lens (UPLFLN100XOP-2, Olympus) and relayed via a tube lens (focal length = 125 mm) onto the imaging sensor. The detection module employs a polarization camera (CS505MUP1, Thorlabs) equipped with an integrated micro-polarizer array, where each 2×2 pixel block contains analyzers oriented at four linear polarization angles (0°, 45°, 90°, 135°). This configuration enables the simultaneous acquisition of four polarization-resolved intensity channels. For volumetric measurements, the sample is axially translated during acquisition using a motorized stepper stage with submicron precision.

The spatial resolution of the system is determined by the optical diffraction limit. Both the condenser and objective lenses have a numerical aperture (NA) of 1.3. Based on the effective spatial-frequency bandwidth in the Fourier domain, the theoretical resolution is estimated to be 0.121 µm laterally and 0.424 µm axially[54].

**Analytic solution of iDTT inverse problem**

In equation (9), the matrix **A** is non-square matrix and may become ill-conditioned at certain spatial-frequency components. To obtain a stable solution, the vector **x** is determined by minimizing the following cost function:

$$C = \| \mathbf{y} - \mathbf{A}\mathbf{x} \|_2^2 + \alpha \| \mathbf{x} \|_2^2 + \beta [3 \operatorname{tr}(\mathbf{F}_r^* \mathbf{F}_r) - \operatorname{tr}(\mathbf{F}_r^*) \operatorname{tr}(\mathbf{F}_r)], \tag{11}$$

where $\operatorname{tr}(\cdot)$ denotes the matrix trace operator. The first term enforces data fidelity, the second term corresponds to Tikhonov regularization to suppress noise amplification from unstable components, and the third term constrains the total birefringence, enforcing noise coherence among the tensor eigenvalues. The parameters $\alpha$ and $\beta$ are frequency-dependent regularization weights, adjusted according to the evaluated noise level at each spatial frequency.

The analytical solution that minimizes Eq. (11) is given by

$$\mathbf{x} = (\mathbf{A}^T\mathbf{A} + \alpha \mathbf{I} + \beta \mathbf{B})^{-1}\mathbf{A}^T\mathbf{y}, \tag{12}$$

where the matrix **B** is a quadratic coefficient matrix of the total birefringence, $\mathbf{x}^T\mathbf{B}\mathbf{x} = 3\operatorname{tr}(\mathbf{F}_r^*\mathbf{F}_r) - \operatorname{tr}(\mathbf{F}_r^*)\operatorname{tr}(\mathbf{F}_r)$.

**Stabilizing inverse problems through conditioning assessment and regularization weight selection**

Although multiple measurements are acquired under diverse illumination and polarization conditions, the inverse problem of reconstructing the dielectric tensor can still be ill-conditioned at specific spatial-frequency components, leading to instability and amplified noise in the reconstructed tensor elements. The conditioning of the problem is quantitatively evaluated from the singular values of the OTF matrix **A** in Eq. (9). The magnitude and distribution of the singular values reflect the degree of stability and effective spatial resolution associated with each singular mode.

To suppress noise amplification arising from poorly conditioned modes, the regularization parameters $\alpha$ and $\beta$ were adaptively weighted according to the conditioning of the system matrix. Specifically, for each spatial-frequency component, $\alpha$ and $\beta$ were scaled in proportion to the inverse square of the smallest non-zero singular value of **A**. This frequency-dependent weighting strategy stabilizes the inversion process while preserving high-frequency features in well-conditioned regions, ensuring robust and spatially uniform tensor reconstruction across the full 3D volume.

**Preparation of colon tissue from Asan medical center**

Formalin-fixed, paraffin-embedded (FFPE) human colon tissue samples were obtained from the Asan Medical Center under approval from the Institutional Review Board (IRB) with a waiver of informed consent (IRB no. 2021-1698). For iDTT imaging, the tissue was sectioned to a thickness of 10 µm using a rotary microtome and subsequently deparaffinized with toluene. The cleared sections were mounted between coverslips using OpticMount™ X as the mounting medium. An adjacent serial section was processed by routine Masson's trichrome (MT) staining and mounted on standard slide glass. The MT-stained section served as a histological reference for assessing the spatial distribution of stromal collagen and radiation-induced fibrosis.

**Grain boundary identification and disorientation mapping**

The local orientation of each voxel's principal axes can be represented by a rotation matrix **g**, which transforms the laboratory coordinate system into the local principal coordinate system. The relative orientation between two neighboring voxels is expressed as

$$\mathbf{g}_{12} = \mathbf{g}_1^{-1} \mathbf{g}_2, \quad (13)$$

where $\mathbf{g}_1$ and $\mathbf{g}_2$ denote the rotation matrices of the two voxels. According to Euler's rotation theorem, any rotation matrix can be represented as a single rotation about a unique axis. The corresponding rotation angle $\theta$ between the two orientations can be computed from the trace of the relative rotation matrix:

$$cos\theta = [\text{tr}(\mathbf{g}_{12} - 1)]/2. \tag{14}$$

Because of the symmetry of optical anisotropy, multiple equivalent rotations can yield the same physical orientation, leading to degeneracy in $\theta$. To resolve this, the disorientation angle, defined as the smallest possible rotation between two equivalent orientations, was determined by evaluating all symmetry-equivalent rotations as

$$\mathbf{g}_{12} = (\mathbf{g}_1 \mathbf{S})^{-1} \mathbf{g}_2 \mathbf{S}, \tag{15}$$

where $\mathbf{S}$ represents the symmetry operators of the optical anisotropy, including axial and combined symmetries.

3D grain-boundary maps and disorientation maps were generated from the disorientation angles between neighboring voxels. Grain boundaries were identified by applying a threshold to disorientation angles exceeding 7° (Fig. 5e). The disorientation map shown in Fig. 5f was computed by comparing disorientation angles within a neighborhood of up to three voxels, with the maximum angle selected to minimize artifacts caused by microcracks or local discontinuities between grains.

**Acknowledgements**

The authors thank Bohyun Ahn and Prof. Pilhan Kim for their assistance with the SHG experiments. This work was supported by the National Research Foundation of Korea (NRF) grant funded by the Korean government (MSIT) (RS-2024-00442348, 2022M3H4A1A02074314), the Korea Institute for Advancement of Technology (KIAT) through the International Cooperative R&D Program (P0028463), the Korean Fund for Regenerative Medicine (KFRM) grant funded by the Ministry of Science and ICT and the Ministry of Health & Welfare (21A0101L1-12), and the Samsung Research Funding Center of Samsung Electronics (SRFC-IT1401-08).

**Data availability**

The datasets generated and analyzed during the current study are available from the corresponding author upon reasonable request.

34